\pdfoutput=1 

\documentclass[3p,times,twocolumn]{elsarticle}

\usepackage{ecrc}
\usepackage{graphicx} 
\usepackage{fixltx2e}

\volume{00}

\firstpage{1}

\journalname{Nuclear Physics B Proceedings Supplement}

\runauth{Alekhin et al.}

\jid{nuphbp}

\jnltitlelogo{Nuclear Physics B Proceedings Supplement}

\usepackage{amssymb}

\usepackage[figuresright]{rotating}

\begin{document}

\begin{frontmatter}



\dochead{}

\title{Nucleon PDF separation with the collider and fixed-target data}


\author{Sergey Alekhin}

\address{Institute for High Energy Physics, Nauki 1, 
   Protvino, Moscow region, 142281 Russia}
\author{Johannes Bl\"umlein, 
Kristin~Lohwasser}\address{
        Deutsches Elektronensynchrotron DESY, Platanenallee 6, D--15738 Zeuthen, Germany}

\author{Lea Michaela~Caminada} 
\address{Physik Institut, Universit\"{a}t Z\"{u}rich,
   Winterthurerstra{\ss}e 190, CH--8057 Z\"{u}rich, Switzerland }

\author{Katerina Lipka, Ringaile Pla\v cakyt\. e} \address{
   Deutsches Elektronensynchrotron DESY, 
   Notkestra{\ss}e 85, D--22607 Hamburg, Germany}

\author{Sven-Olaf Moch}\address{
II. Institut f\"ur Theoretische Physik Universit\"at Hamburg,
    Luruper Chaussee 149, D-22761 Hamburg, Germany}

\author{Roberto Petti}\address{
    Department of Physics and Astronomy, University of South Carolina 
   712 Main Street, Columbia, SC 29208, USA}

\begin{abstract}
We consider the impact of the recent data obtained by the LHC,
Tevatron, and fixed-target experiments on the nucleon quark distributions 
 with a particular focus on disentangling different quark
species. An improved determination of the poorly known strange sea
distribution is obtained due to including data from the 
neutrino-induced deep-inelastic scattering experiments NOMAD and CHORUS. The impact of 
the associated $(W + c)$ production data by CMS and ATLAS on the strange 
sea determination is also studied 
and a comparison with earlier results based on the 
collider data is discussed. Finally, the recent LHC and Tevatron data on the 
charged lepton asymmetry are compared to the NNLO ABM predictions and 
the potential of this input in improving the non-strange sea distributions is 
evaluated.
\end{abstract}

\begin{keyword}
Parton distributions \sep strangeness \sep hadron colliders 
\end{keyword}

\end{frontmatter}


\section{Introduction} 
The nucleon quark distributions are important ingredients of the 
high-energy hadron collision phenomenology, particularly at large Bjorken $x$, 
since these are the quarks, which dominate the partonic content of the nucleon 
for these kinematics. The fixed-target inclusive deep-inelastic scattering 
(DIS) data, which play a central role in the determination of the parton 
distribution functions (PDFs), provide a constraint only on the linear 
combinations of the quark distribution.  
\begin{figure}[h]
  \centering
  \includegraphics[width=0.45\textwidth]{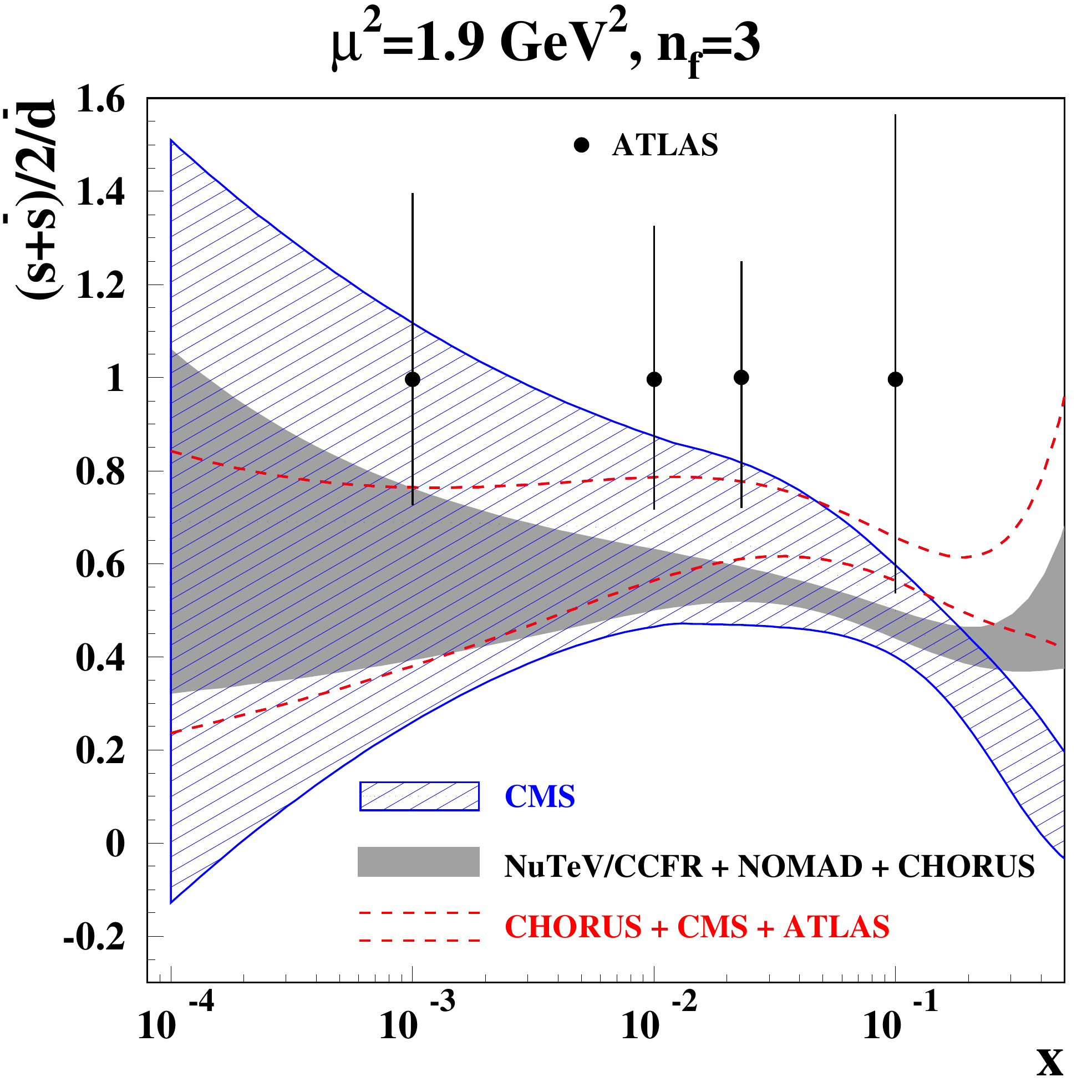}
  \caption{\small    The $1\sigma$ band for the 
    strange sea suppression factor $r_s=(s+\bar{s})/2/\bar{d}$ as a function of the 
    Bjorken $x$
      obtained in the variant of the present analysis based on  
    a combination of the data by NuTeV/CCFR~\cite{Goncharov:2001qe},
 NOMAD~\cite{Samoylov:2013xoa}, and 
CHORUS~\cite{KayisTopaksu:2011mx} (shaded area)
in comparison with the results obtained by CMS~\cite{Chatrchyan:2013mza}
(hatched area), in the ATLAS $epWZ$-fit~\cite{Aad:2014xca,Aad:2012sb}
(circles),
and with the variant of present analysis based on combination of the data by
CHORUS~\cite{KayisTopaksu:2011mx}, 
CMS~\cite{Chatrchyan:2013uja}, and ATLAS~\cite{Aad:2014xca} (dashes) at 
different values of $x$.
All quantities refer to the factorization scale $\mu^2=1.9~{\rm GeV}^2$.} 
  \label{fig:ssup}
\end{figure}
\begin{figure}[h]
  \centering
  \includegraphics[width=0.45\textwidth]{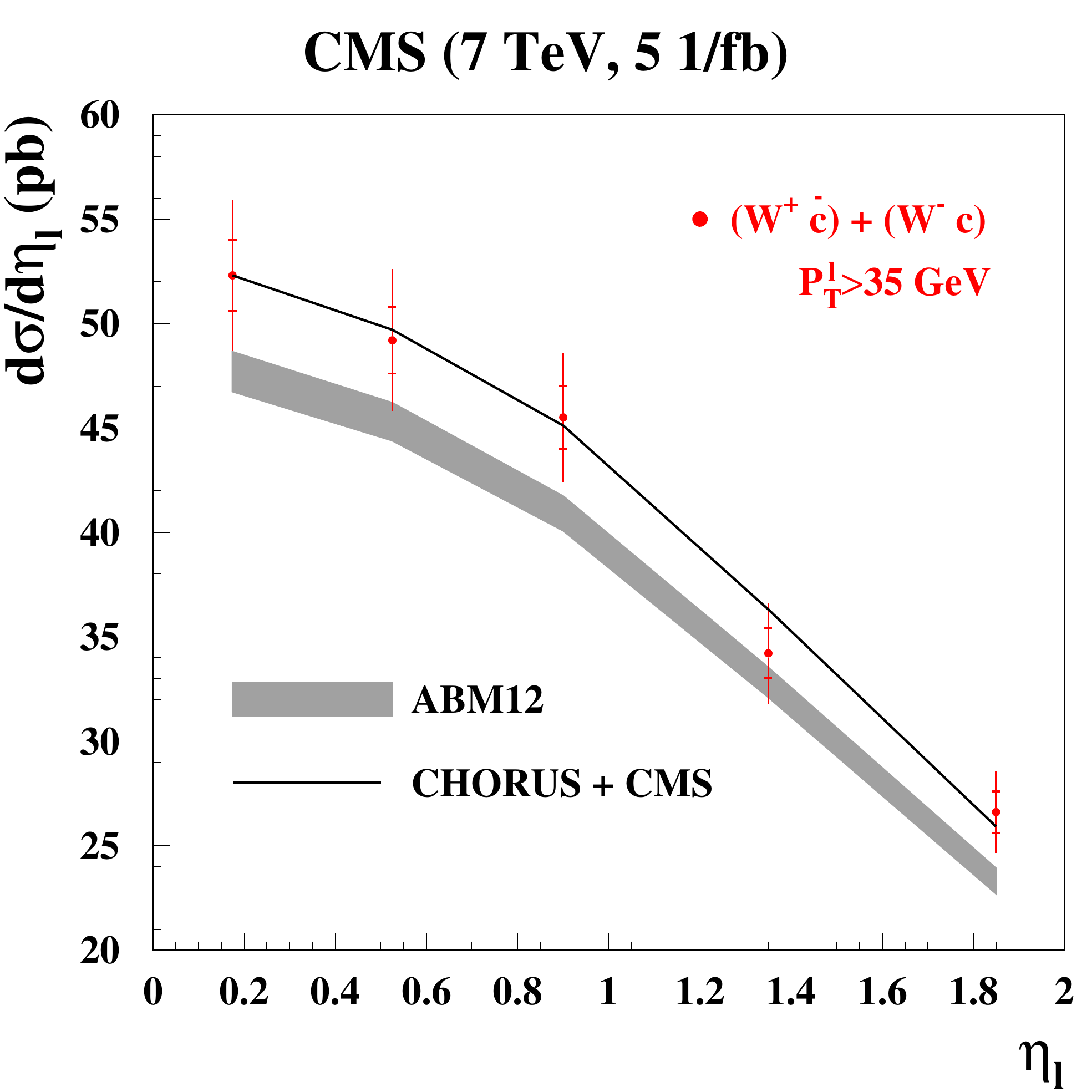}
  \caption{\small
  The CMS data on ($W + c$) production~\cite{Chatrchyan:2013uja} 
in comparison with 
the ABM12 predictions (shaded area) and the variant of ABM12 fit~\cite{Alekhin:2013nda} with  
only the CHORUS charm~\cite{KayisTopaksu:2011mx} and  the 
CMS $(W + c)$~\cite{Chatrchyan:2013uja} data used to constrain the strange sea. 
}
    \label{fig:cmsc}
\end{figure}
Therefore, the global PDF fits 
are commonly based on a combination of the DIS data with ones for the 
Drell-Yan (DY) process, which yield a supplementary constraint on the PDFs. 
At that the strange sea distribution can be independently determined from 
the data on the $c$-quark production in neutrino-nucleon DIS probing the 
strange sea by the charged-current transition $s\rightarrow c$. If the
$c$-quark produced is detected by its semi-leptonic decay into a muon,
this process can be triggered by two muons in the final states. This 
signature was used by the NuTeV/CCFR 
experiments~\cite{Goncharov:2001qe} at 
Tevatron and the NOMAD experiment~\cite{Samoylov:2013xoa} at CERN.
Alternatively, the events with a $c$-quark in the
final state can be selected 
by the hadronic decays of charmed hadrons using the emulsion technique,
like in the CHORUS experiment~\cite{KayisTopaksu:2011mx}.
The NuTeV/CCFR data~\cite{Goncharov:2001qe}
have been employed to constrain the strange sea in the ABM PDF 
fit~\cite{Alekhin:2008mb, Alekhin:2012ig, Alekhin:2013nda}. 
In the present proceedings we describe how the recent 
NOMAD~\cite{Samoylov:2013xoa} and 
CHORUS~\cite{KayisTopaksu:2011mx} data 
can further improve the strange sea determination. We also discuss 
the associated ($W + c$) production in the proton-proton collisions 
measured by the CMS~\cite{Chatrchyan:2013uja} 
and ATLAS~\cite{Aad:2014xca} experiments at the LHC. Similarly to the 
case of neutrino-induced DIS, this process 
receives essential contributions from 
the charged-current transition $s\rightarrow c$. 
Therefore, it can be also used to pin down the strangeness contribution to the nucleon. 
Finally, we check the recent data on the charged lepton asymmetry $A_l$ by 
the CMS~\cite{Chatrchyan:2013mza} and D0~\cite{Abazov:2013rja} experiments measured in 
(anti-)proton-proton collisions  
w.r.t.~the ABM12 PDFs~\cite{Alekhin:2013nda}. 
Since the $A_l$ measurement gives a constraint allowing to
separate the $u$- and $d$-quark distributions this comparison  
demonstrates a further trend in determination of the 
non-strange quark PDFs.
\begin{figure*}[t]
  \centering
  \includegraphics[width=0.95\textwidth,height=8.5cm]{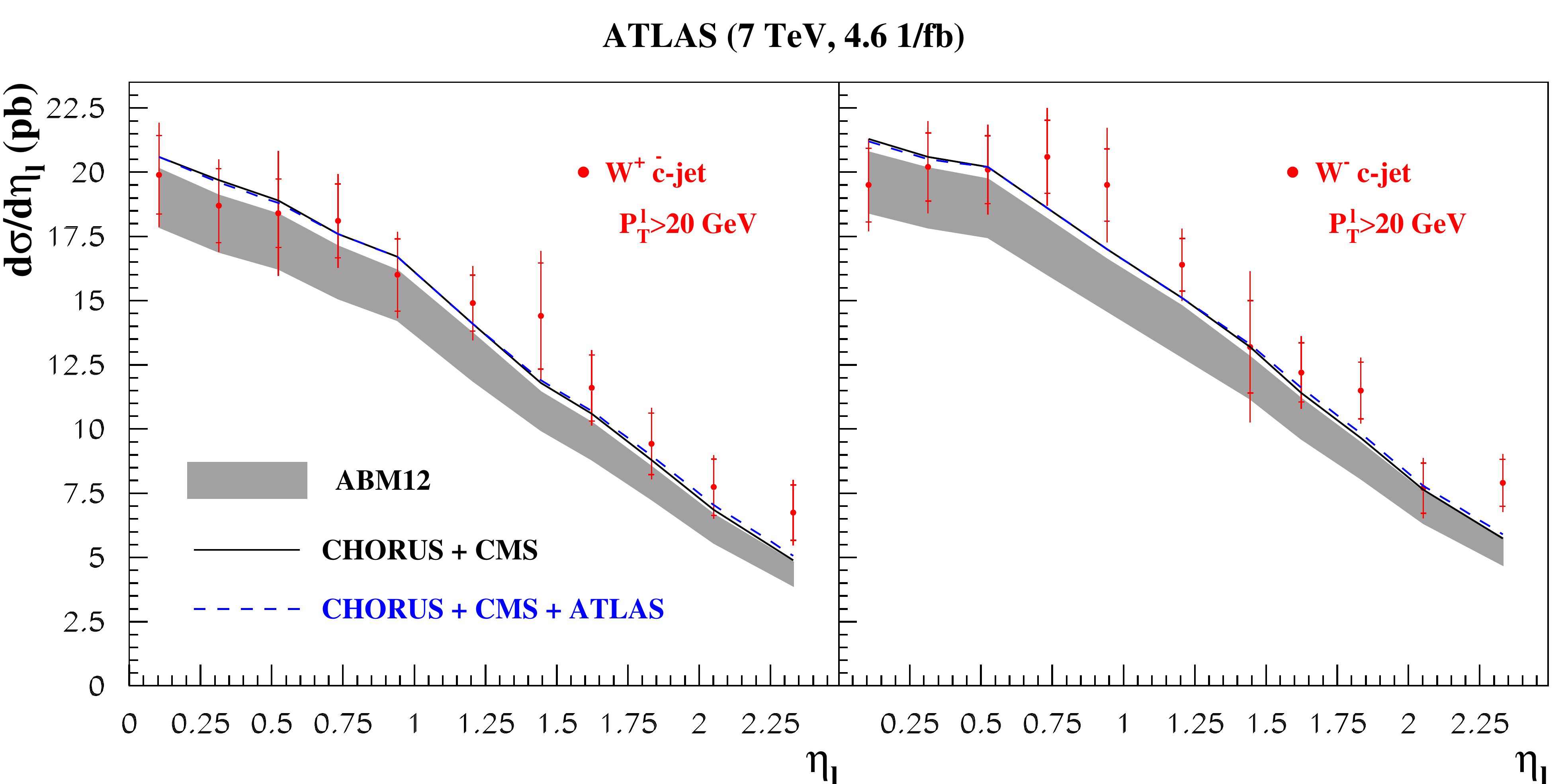}
  \caption{\small
The same as in Fig.~{\protect \ref{fig:cmsc}}
 for the ATLAS $(W^+ + \bar c) $ (left)
and $(W^- + c) $ (right) data~\cite{Aad:2014xca}. 
The variant of the ABM12 fit~\cite{Alekhin:2013nda} with  
only CHORUS charm data~\cite{KayisTopaksu:2011mx} in combination 
with the CMS and ATLAS $(W + c)$
ones~\cite{Chatrchyan:2013uja,Aad:2014xca} used to constrain the strange sea
is displayed by dashes.
}
    \label{fig:atlasc}
\end{figure*}

\section{Strange sea improvement} 
The NOMAD experiment collected about 15K dimuon events in the neutrino-nucleon 
DIS~\cite{Samoylov:2013xoa} exceeding the total statistics of earlier 
NuTeV/CCFR samples~\cite{Goncharov:2001qe} by a factor of 1.5.  
Furthermore the NOMAD beam energy is essentially lower than 
the NuTeV/CCFR one that allows to probe the strange sea in the 
kinematic region inaccessible for NuTeV/CCFR, up to $x\sim 0.5$.
The strange sea distribution can be extracted from the dimuon data provided 
the semi-leptonic branching ratio of the produced charmed hadrons
is known. Meanwhile, their
production rate essentially depends on the beam energy $E_\nu$ therefore
an effective semi-leptonic branching ratio
$B_\mu(E_\nu)=\sum _h r_h(E_\nu) B_\mu^h$
where $r_h(E_\nu)$ is the relative production rate of hadron $h$ and 
$B_\mu^h$ is its semi-leptonic branching ratio, also depends on $E_\nu$. 
We take into account this dependence employing an empirical parameterization 
of Ref.~\cite{Samoylov:2013xoa}
\begin{equation}
B_\mu(E_\nu)=\frac{B_\mu^{(0)}}{1+B_\mu^{(1)}/E_\nu}
\label{eq:bmu}
\end{equation}
with the parameters $B_\mu^{(0,1)}$ constrained by the existing data on 
$B_\mu$~\cite{Ushida:1988rt,KayisTopaksu:2005je}. 
The values of $B_\mu^{(0)}=0.0933(25)$ and  $B_\mu^{(1)}=5.6\pm1.1$
obtained in the present analysis are in 
agreement with the ones obtained by NOMAD~\cite{Samoylov:2013xoa}. 
With these values the parameterization Eq.~(\ref{eq:bmu}) flattens
in the beam energy range of NuTeV/CCFR and is consistent with the 
value of $B_\mu$ used in the earlier ABM12 fit~\cite{Alekhin:2013nda}.

The CHORUS experiment collected 2013 events 
with the charmed hadrons produced in the neutrino-nucleon DIS using 
the emulsion detection technique~\cite{KayisTopaksu:2011mx}. This is 
much smaller than the statistics of the dimuon experiments. However, due to 
the charm-decay vertex reconstruction, the CHORUS data are not sensitive to 
the corresponding branching ratios and provide in this way a complementary 
constraint on the strange sea. 

The strange sea distribution obtained in the updated version of the ABM12 
fit where the NOMAD and CHORUS data included is given 
in Fig.~\ref{fig:ssup} as a ratio $r_s=(s+\bar{s})/2/\bar{d}$ 
quantifying the suppression of the strange sea w.r.t.~the non-strange one.  
\begin{figure}[h]
  \centering
  \includegraphics[width=0.45\textwidth]{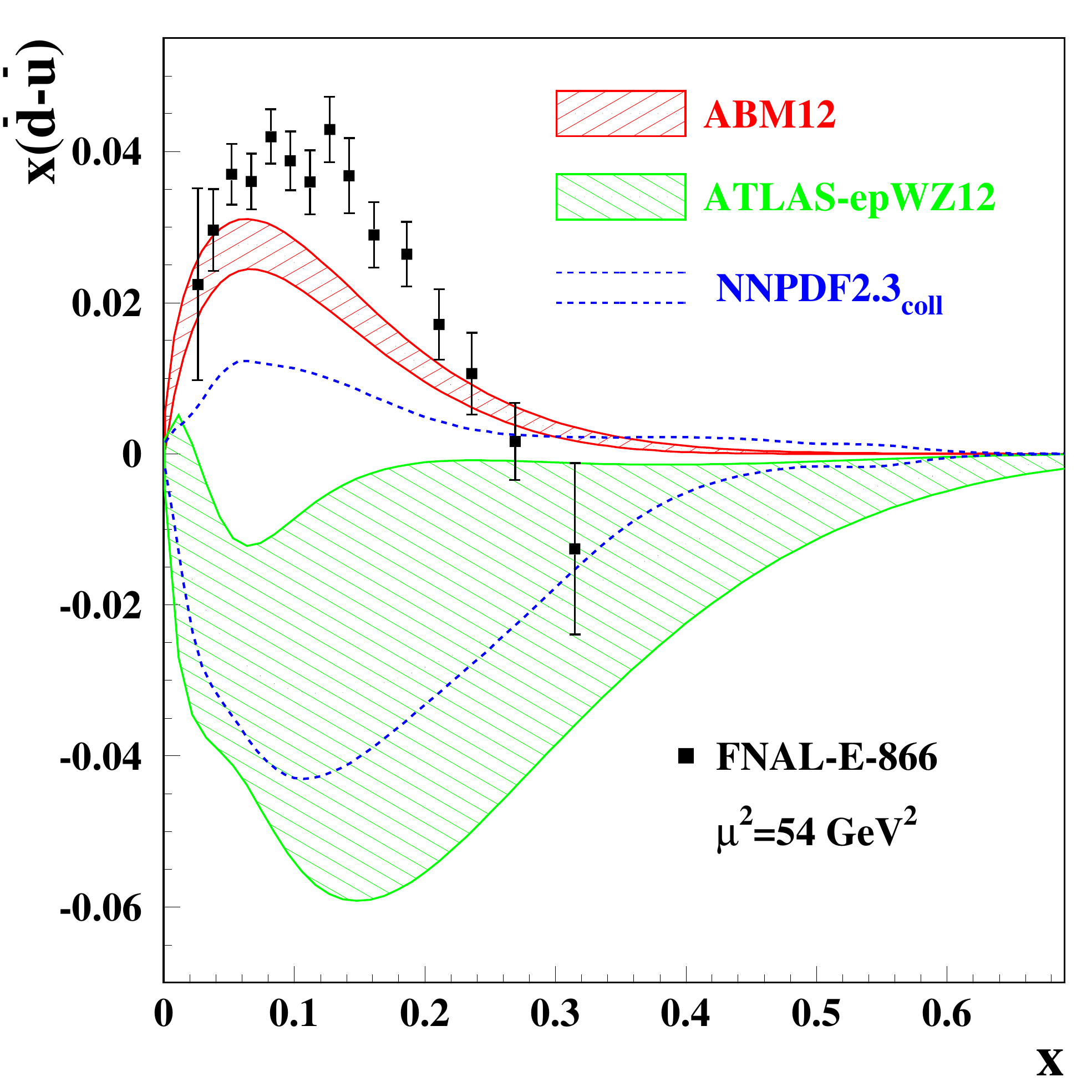}
  \caption{\small
    The $1\sigma$ band for 
    the iso-spin asymmetry of the sea $x(\bar{d}-\bar{u})$ at the scale of
    $\mu^2=54~{\rm GeV}^2$ as a function of the Bjorken $x$ obtained in the ABM12 fit~\cite{Alekhin:2013nda} 
    (right-tilted hatch) in comparison with the corresponding ones obtained in 
    the ATLAS~\cite{Aad:2012sb} (left-tilted hatch) and 
the NNPDF~\cite{Ball:2012cx} (dashes) analyses using only the LHC and HERA 
collider data. The values of $x(\bar{d}-\bar{u})$ extracted from the 
FNAL-E-866 data~\cite{Towell:2001nh} within the Born approximation are  
displayed as full circles.  
}
    \label{fig:udm}
\end{figure}
It is roughly constant in a wide range of $x$ and consistent with the value 
obtained earlier in our analysis of the NuTeV/CCFR dimuon data. 
However, the error in the updated determination of the strange sea at $x\gtrsim 0.01$ 
is improved up to a factor 2, mainly due to the impact of the NOMAD 
data~\cite{Samoylov:2013xoa} 
(cf. more details in Ref.~\cite{Alekhin:2014sya}).
The CHORUS
data~\cite{KayisTopaksu:2011mx} somewhat overshoot the 
fit and therefore pull the strange sea up. 

The same is valid 
for the CMS and ATLAS data on $(W + c)$ 
production~\cite{Chatrchyan:2013uja,Aad:2014xca}, 
cf. Figs.~\ref{fig:cmsc}, \ref{fig:atlasc},
although in this case 
the tension can be at least partially explained by the impact of the 
still missing NNLO QCD corrections to this process. 
Note, the ATLAS data are in a good agreement with the predictions based on 
a variant of the ABM12 fit 
with only CHORUS and CMS data are used to constrain the 
strange sea, while the CMS ones are also described very well in this 
variant of fit. This demonstrates a good agreement between these
three samples, which all together prefer enhanced strange sea. 
To check the upper margin of $r_s$ allowed by this data subset
we consider the variant of the 
fit with the NuTeV/CCFR and NOMAD data dropped and only the CHORUS, CMS, 
and ATLAS data used to constrain the strange sea.
\begin{figure}[h]
  \centering
  \includegraphics[width=0.45\textwidth]{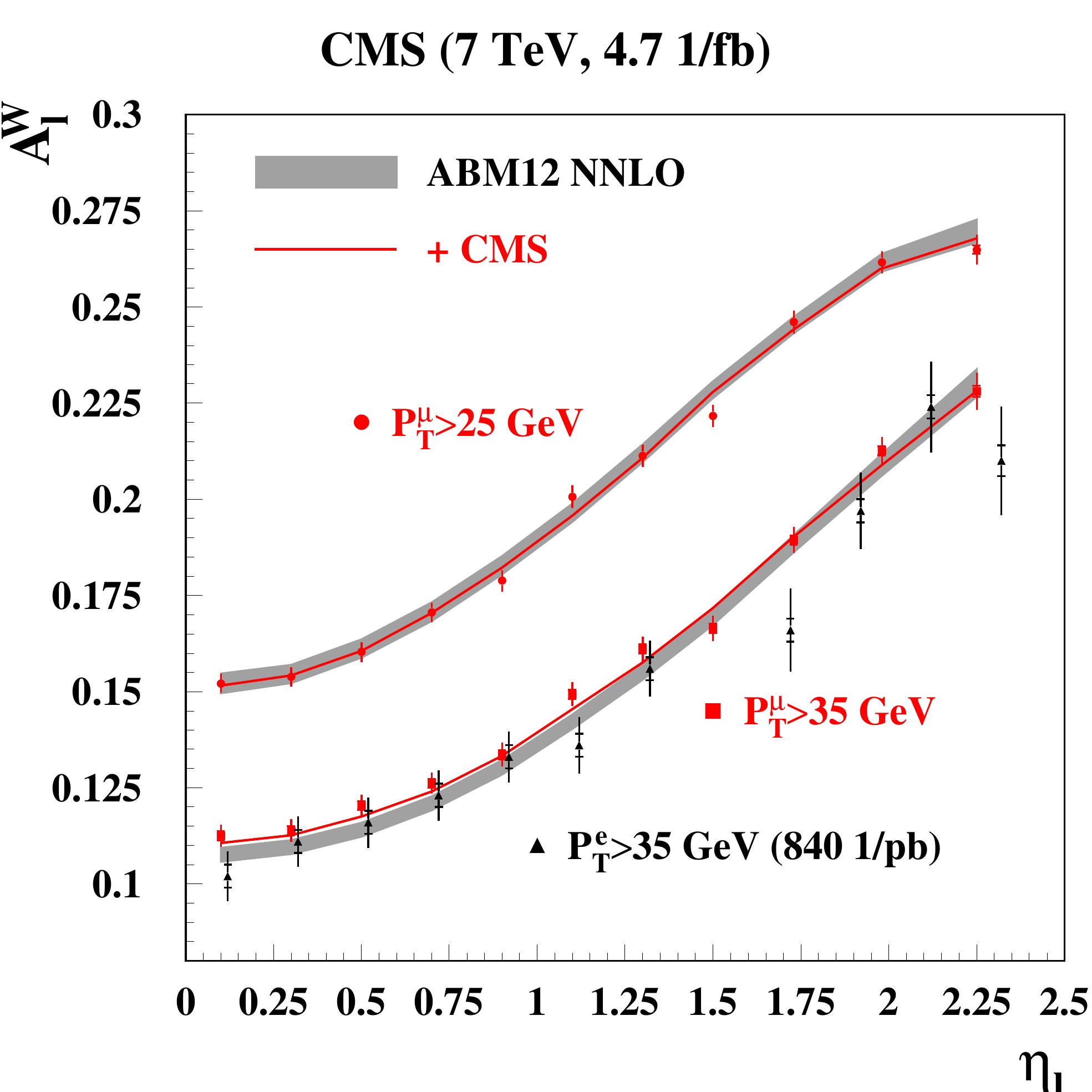}
  \caption{\small
The CMS data on the muon asymmetry~\cite{Chatrchyan:2013mza} 
with the cuts of $P_T^{~\mu}>25~{\rm GeV}$ (circles) 
and $P_T^{~\mu}>35~{\rm GeV}$ (squares)
imposed on the transverse muon momentum in comparison with the 
NNLO ABM12 prediction error band due to the PDF uncertainties (shaded area)
and the earlier CMS data on the electron 
asymmetry~\cite{Chatrchyan:2012xt} with the cut 
of $P_T^{~\mu}>35~{\rm GeV}$ (triangles). The curves display the variant 
of ABM12 fit~\cite{Alekhin:2013nda} with the CMS data of Ref.~\cite{Chatrchyan:2013mza} included. 
 }
    \label{fig:cms}
\end{figure}
In this case the value of 
$r_s$ at $x\sim 0.1$ is enhanced as compared to the determination based on the 
combination of the NuTeV/CCFR, NOMAD, and CHORUS data. However the difference 
is within $2\sigma$ and it does not exceed 20\% at maximum. 

At the same time, the strange sea obtained in the ATLAS PDF 
analysis~\cite{Aad:2012sb}
is comparable with the non-strange one, cf. Fig.~\ref{fig:ssup}.
The analysis of Ref.~~\cite{Aad:2012sb} is based on  
the ATLAS data on the inclusive production of the $W$- and $Z$-bosons 
taken in combination with the HERA inclusive DIS data, 
while all fixed-target DIS and DY data are not considered. However, the 
result of the ATLAS analysis is in disagreement 
with the data by the FNAL-E-866 experiment on 
the ratio of the proton and deuteron 
DY-process cross sections~\cite{Towell:2001nh}, which are commonly used to
constrain the sea iso-spin asymmetry 
$x(\bar d- \bar u)$ in global PDF fits. Indeed, the value  
of $x(\bar d- \bar u)$ extracted from the FNAL-E-866 data in the Born 
approximation is positive for the whole range of $x$ and in the 
ATLAS analysis~\cite{Aad:2012sb} negative values are preferred. 
This means, the enhancement of the strange sea observed by ATLAS is obtained
at the expense of 
a suppression of the $d$-quark PDF. Note, that 
negative values of $x(\bar d- \bar u)$ were also 
observed in the NNPDF fit~\cite{Ball:2012cx} 
based on collider data only and a corresponding strange sea enhancement 
appears there, too. Meanwhile, the discrepancy between the 
ATLAS results and the FNAL-E-866 data is not dramatic and amounts to about $2\sigma$ only.
Moreover, the ATLAS result on $r_s$ is in agreement with ours within the errors, cf. Fig.~\ref{fig:ssup}.
On the other hand, the ATLAS data on the $W$- and $Z$-boson 
production~\cite{Aad:2011dm} are 
well described by the PDF set with the suppressed strange sea. 
In particular, in the updated variant of the ABM12 fit with the NOMAD and 
CHORUS data included we get a value of $\chi^2=35$ for 30 data points in the 
combined ATLAS $W$- and $Z$-data set. In summary, these findings  
rather point at an insufficient statistical potential of the data used in the 
ATLAS analysis of Ref.~\cite{Aad:2012sb}  
in disentangling the quark PDFs than at a real
 strange sea enhancement.
In this context it is also worth noting that in the 
analysis of Ref.~\cite{Chatrchyan:2013mza} based on the combination of 
the 
CMS and HERA data the strange sea suppression at $x\gtrsim 0.01$ was observed,
cf. Fig.~\ref{fig:ssup}.
\begin{figure}[t]
  \centering
  \includegraphics[width=0.45\textwidth]{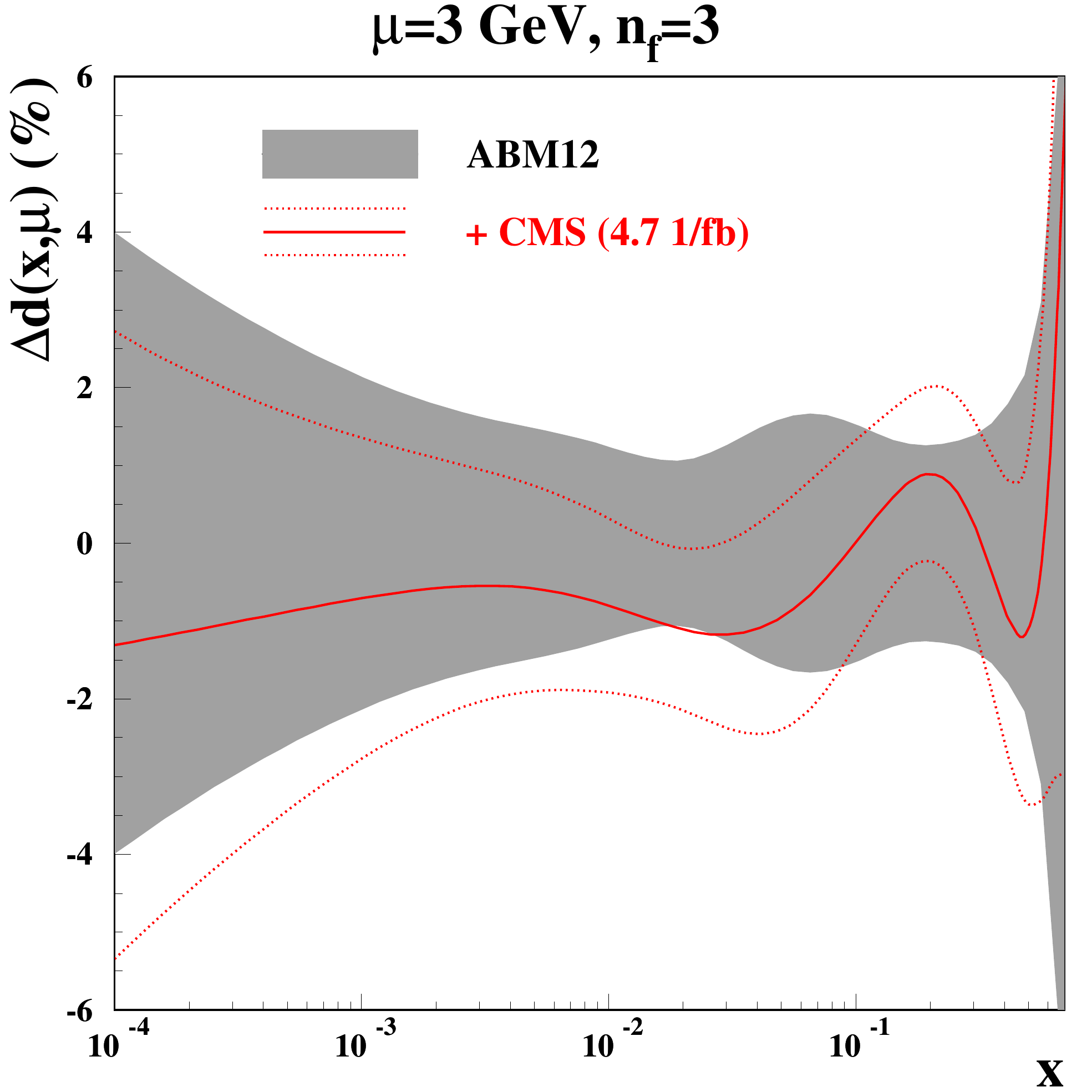}
  \caption{\small
The $1\sigma$ fractional error band for the ABM12 $d$-quark 
distribution (shaded area) in comparison with the relative variation
of its central value due 
to adding the CMS muon asymmetry data~\cite{Chatrchyan:2013mza} (solid curve). 
The dotted curves correspond to the $d$-quark distribution 
uncertainty in this variant of the ABM12 fit~\cite{Alekhin:2013nda}. 
 }
    \label{fig:cmsd}
\end{figure}

\section {Constraints on the non-strange quark PDFs} 
Recent data on the muon asymmetry $A_\mu$ obtained by 
CMS~\cite{Chatrchyan:2013mza} for the sample with an integrated luminosity 
of 4.7 1/fb are compared with the predictions based on the NNLO ABM12 
PDFs~\cite{Alekhin:2013nda} in Fig.\ref{fig:cms}.
The predicted central values and their uncertainties due to PDFs are 
computed with {\tt DYNNLO}~(version 1.3)~\cite{Catani:2009sm} and   
{\tt FEWZ}~(version 3.1~\cite{Li:2012wna}), respectively, because the former 
code shows a 
better numerical convergence, while the latter allows for a parallel 
computation with different PDF set members. 
The agreement between data and the predictions is in general quite good.
For the data sample with $P_T^{~\mu}>35~{\rm GeV}$ the predictions do somewhat
undershoot the data at the muon pseudo-rapidity $\eta_\mu\lesssim 1$
since the ABM12 PDF were tuned to the earlier CMS 
data on the electron asymmetry~\cite{Chatrchyan:2012xt},
which go lower than the muon ones, cf. Fig.~\ref{fig:cms}. However, this 
tension is within the 1$\sigma$ PDF uncertainties. Furthermore, 
for the data sample with 
$P_T^{~\mu}>25~{\rm GeV}$ a tension does not show up. 
Therefore the ABM12 fit does not change significantly due to inclusion of 
the CMS data on the electron asymmetry. 
The values of $\chi^2$ obtained in 
the variants of this fit with the
data at $P_T^{~\mu}>25~{\rm GeV}$ and 
$P_T^{~\mu}>35~{\rm GeV}$ included are 16 and 10, respectively, for 11 data
points. The former value is somewhat worse than the ideal one due to 
the data fluctuations at $\eta_\mu \sim 1.5$ going beyond the experimental 
errors quoted. The change in the ABM12 PDFs due to the CMS muon asymmetry data
is within $1\sigma$ with the most significant impact observed for the 
$d$-quark distribution. The uncertainty in the latter at $x\sim 0.1$ is 
also somewhat improved, cf. Fig.~\ref{fig:cmsd}. 
\begin{figure*}[t]
  \centering
  \includegraphics[width=0.95\textwidth,height=7.5cm]{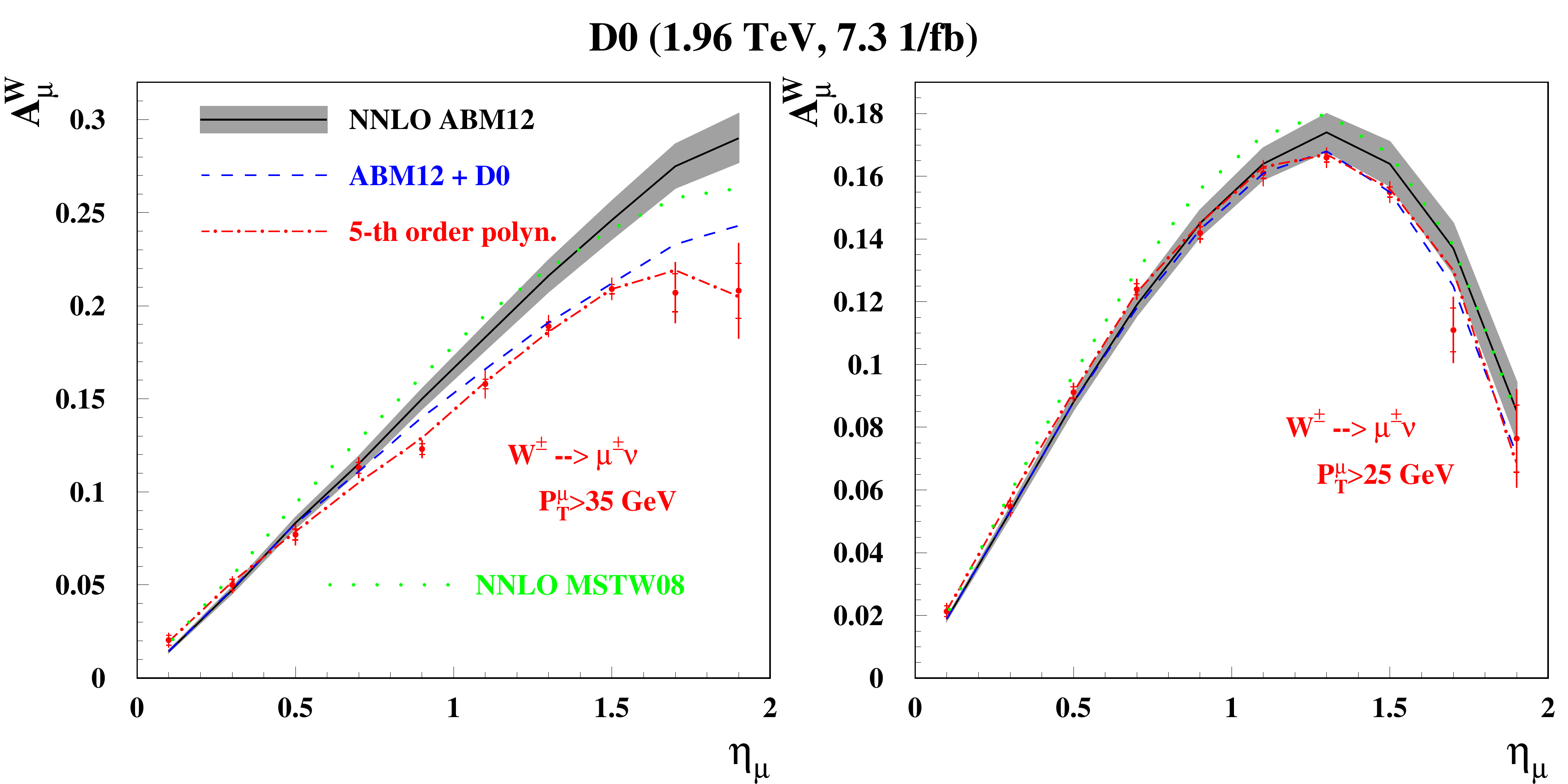}
  \caption{\small
The muon asymmetry data obtained by the D0 
collaboration~\cite{Abazov:2013rja} for integral 
luminosity of 7.3 1/fb with the cuts on the muon transverse momentum 
 $P_T^{~\mu}>35~{\rm GeV}$ (left panel) and  $P_T^{~\mu}>25~{\rm GeV}$ 
(right panel) in comparison with the NNLO predictions based on the  
ABM12 PDFs (solid curves: central values, shaded area: the uncertainties
 due to PDFs) and the central values for the 
MSTW08~\cite{Martin:2009iq} ones (dots). 
The variant of the ABM12 fit~\cite{Alekhin:2013nda} including the D0 data~\cite{Abazov:2013rja} and 
the model-independent polynomial fit to the D0 data only are displayed
by dashes and dotted-dashes, respectively. 
 }
    \label{fig:d0}
\end{figure*}

In contrast with the CMS muon asymmetry and the related LHC data in general, 
the Tevatron data on the 
charged lepton asymmetry do not demonstrate a good  agreement with the ABM12 
predictions. In particular, recent D0 results on the muon asymmetry~\cite{Abazov:2013rja}
obtained with the 
integral luminosity of 7.3 1/fb significantly overshoot the ABM12 predictions
at $\eta_\mu \gtrsim 1$, cf. Fig.~\ref{fig:d0}. A discrepancy is especially 
pronounced for the sample with the cut of $P_T^{~\mu}>35~{\rm GeV}$.
In this case the shape of the ABM12 predictions differ from the 
data even qualitatively. The agreement is improved in case the D0 data are 
added to the ABM12 fit. However, even in this case the profile of data 
at $P_T^{~\mu}>35~{\rm GeV}$ is not entirely
reproduced and the value of $\chi^2$ is 40 for 10 data points. A better 
value of $\chi^2=14$ can be achieved in a model-independent fit with a 
5-th order polynomial. However, in this case the fitted curve exhibits a 
step at $\eta_\mu \sim 1$, which evidently cannot be provided with a smooth 
PDF shape. The same step, although less pronounced, appears in the profile of 
the D0 sample 
with $P_T^{~\mu}>25~{\rm GeV}$, cf. Fig.~\ref{fig:d0}. It is worth 
noting that the MSTW08 PDFs being tuned to the earlier Tevatron data also 
demonstrate poor agreement with the D0 data of Ref.~\cite{Abazov:2013rja}. 
This may point to a disagreement with the earlier Tevatron data. These issues  
should evidently be clarified before the recent D0 data 
can be included into our PDF fit. 

\section {Conclusion} 
In conclusion, we find that the ABM strange sea distribution can be 
essentially improved at $x\gtrsim 0.01$ by adding to the fit 
recent NOMAD and CHORUS data on the charm production in the 
DIS neutrino-nucleon scattering. The updated PDFs are in 
agreement with the ABM12 ones although the CHORUS data 
obtained on the emulsion target
demonstrate a particular trend w.r.t. the results of dimuon experiments
NuTev, CCFR, and NOMAD
pulling strange sea somewhat up. The CMS and ATLAS data
on the associated $(W + c)$ production in the proton-proton collisions, 
which are also sensitive to the strangeness content of nucleon, 
overshoot the ABM predictions, however, statistical significance of the 
discrepancy is inessential in view of the large data uncertainties
and foreseen impact of the NNLO corrections to this process. 
From the variant of the ABM fit based on the CHORUS, CMS, and ATLAS data 
preferring enhanced strange sea we estimate  
an upper margin of such enhancement as 20\%.

The recent CMS high-statistics
data on the muon asymmetry are in a good agreement with 
the ABM12 NNLO predictions and allow further improvement in 
disentangling the $u$- and $d$-quark distributions. As a result, the 
uncertainty in ABM $d$-quark distribution at $x\sim 0.1$ can be reduced 
by adding the CMS  sample to the fit. At the same time, the D0 data on 
the muon asymmetry, which can potentially further 
improve the quark PDF uncertainties at bigger values of $x$,
demonstrate poor agreement with the ABM12 predictions and cannot be employed 
in our analysis before clarification of these discrepancies.

{\bf Acknowledgments.} We would to thank P.~Marquard for careful reading of the text and valuable 
comments. This work has been supported in part 
by Helmholtz Gemeinschaft under contract VH-HA-101 ({\it Alliance Physics at the Terascale}), 
DFG Sonderforschungsbereich/Transregio~9 and by the European Commission through contract      
PITN-GA-2010-264564 ({\it LHCPhenoNet}). We thank the Galileo Galilei Institute 
for Theoretical Physica for the hospitality and the INFN for partial support 
during the completion of this work.

\end{document}